\documentclass{phbauth}
\usepackage{graphicx,epsfig,float}

\begin{document}

\begin{frontmatter}

% Use lower case letters in the title.
\title{From Mott insulator to overdoped superconductor:\\ Evolution of
the electronic structure of cuprates studied by ARPES}

\author{A. Damascelli\thanksref{thank1}},
\author{D.H. Lu},
\author{Z.-X. Shen
}
\address{Department of Physics, Applied Physics and Stanford Synchrotron
        Radiation Laboratory\\ Stanford University, Stanford, CA\,94305, USA}

\thanks[thank1]{Corresponding author. E-mail: damascel@stanford.edu}

\begin{abstract}
We review our angle-resolved photoemission spectroscopy (ARPES)
results on different layered oxide superconductors, and on their
insulating parent compounds. The low energy excitations are
discussed with emphasis on some of the most recent issues, such as
the Fermi surface and remnant Fermi surface, the pseudogap and
$d$-wave-like dispersion, and lastly the signatures in the ARPES
spectra of multiple electronic components, many body effects, and
the superfluid density. We will focus on systematic changes in the
electronic structure which may be relevant to the development of a
comprehensive picture for the evolution from Mott insulators to
overdoped superconductors.
\end{abstract}

\end{frontmatter}

\section{Introduction}

\begin{figure}[t!]
\centerline{\epsfig{figure=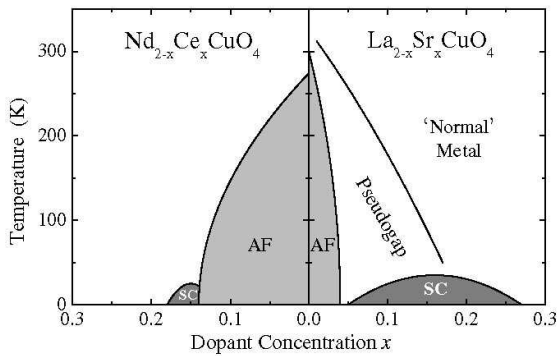,width=0.97\linewidth,clip=}}
\vspace{-.2cm} \caption{Phase diagram of n and p-type
superconductors.}\label{phased}\end{figure}

Following their remarkable discovery in 1986 \cite{bednorz,wu},
high-temperature superconductors (HTSCs) have attracted great
interest due to their scientific significance and enormous
potential for applications. The latter is obviously related to the
high transition temperature shown by these compounds ($T_c$ can be
as high as 134 K in HgBa$_2$Ca$_2$Cu$_3$O$_{8-\delta}$ at
atmospheric pressure \cite{schilling}). Their scientific
importance stems from the fact that the HTSCs highlight a major
intellectual crisis in the quantum-theory of solids which, in the
form of one-electron band theory, has been very successful in
describing good metals and semiconductors (like Cu and Si,
respectively) but has proven to be inadequate for strongly
correlated electron systems. The failure of the single-particle
picture and the main conceptual issues involved in the study of
HTSCs can be best illustrated starting from the phenomenological
phase diagram of n and p-type HTSCs [represented by
Nd$_{2-x}$Ce$_x$CuO$_4$ (NCCO) and La$_{2-x}$Sr$_x$CuO$_4$ (LSCO),
respectively, in Fig.\,\ref{phased}].

The $T^2$ dependence of the resistivity observed in the overdoped
metallic regime is taken as evidence for Fermi liquid (FL)
behavior. On the other hand, the applicability of FL theory (which
describes electronic excitations in terms of a weakly interacting
gas of {\it quasiparticles}) to the `normal' metallic state of
HTSCs is questionable, because many properties do not follow
canonical FL behavior. Most dramatic becomes the breakdown of FL
theory and of the single particle picture upon approaching the
undoped line of the phase diagram ($x\!=\!0$), where we find the
antiferromagnetic (AF) Mott insulator. Mott-Hubbard insulators,
because of the odd number of electrons per unit cell, are systems
erroneously predicted by band theory to be paramagnetic metals
(with a partially filled {\it d}-band in the case of transition
metal oxides) \cite{mott,hubbard,pwa}. The reason for this failure
lies in the on-site electron-electron repulsion $U$ which is much
larger than the bandwidth $W$. As a consequence, charge
fluctuations are suppressed and these compounds are rather good
insulators at all temperatures, with an optical gap $U$ of a few
eV between the lower and upper Hubbard bands (LHB and UHB). As a
matter of fact, in the cuprates the Cu-O charge-transfer energy
$\Delta$ is smaller than the on-site Coulomb repulsion $U$ (see
Fig.\,\ref{mott}a), which characterizes these compounds more
precisely as charge-transfer insulators \cite{zsa}. However,
because the first electron-removal state corresponds to the
O-derived Zhang-Rice singlet band (ZRB), the cuprates are thought
to be equivalent to an effective Mott-Hubbard system with the ZRB
playing the role of the LHB, and an in-plane Cu-derived band as
the UHB \cite{zhang}. These states are separated by an effective
Mott gap\,$\sim\!\Delta$.

Therefore, many aspects of the physics of the cuprates are
believed to be captured by the single-band Hubbard model
\cite{rvb}. This contains a kinetic-energy term proportional to
the nearest neighbor ({\it nn}) hopping amplitude $t$, and an {\it
ad hoc} Hubbard $U$ term which accounts for electronic
correlations: large Coulomb repulsion $U$ favors electron
localization and results in `frustration' of the kinetic energy.
In the strong coupling limit ($U\!\gg\!t$) at half filling
($x\!=\!0$, i.e., one electron per Cu site), the AF state
\cite{pwa1} results from the fact that, when $nn$ spins are
antiparallel to each other, the electrons gain kinetic energy by
undergoing virtual hopping to the neighboring sites (because of
the Pauli principle hopping is forbidden for parallel spins). By
projecting out the doubly occupied states at large $U$
\cite{dagotto},  the low-lying excitations of the 1/2-filled
Hubbard model are described by the $t$-$J$ Hamiltonian
($J\!\sim\!t^2\!/U$). Away from half filling, the $t$-$J$ model
describes the so called `doped AF', i.e., a system of interacting
spins and mobile holes. The latter acquire a `magnetic dressing'
because they are perturbing the correlations of the spin
background that they move through.

The challenge in the investigation of the electronic properties of
the cuprates is to sort out the basic phenomenology that can test
the relevance of many-body models in describing the low lying
excitations, in both the insulator and the doped metal. At the
same time, it is also prudent to consider the influence of other
degrees of freedom on the physical properties of these complex
materials. For instance, phonon modes and lattice distortion can
play a very important role, in particular when they couple to
potential instabilities of the charge and/or spin degrees of
freedom. Nevertheless, in order to address the scope of the
current approach in the quantum theory of solids and the validity
of the proposed models, a detailed comparison with experiments
that probe the electronic properties and the nature of the
elementary excitations is required. In this context, ARPES has
played a major role because it is the most direct method of
studying the electronic structure of solids \cite{zx}.

In this paper, we review our recent ARPES results on the cuprates.
As the doping evolution allows a critical comparison between
theory and experiment, we will discuss ARPES data on the HTSCs and
their insulating parent compounds, focusing on systematic changes
in the electronic structure that may be relevant to the
development of a comprehensive picture for the evolution from Mott
insulators to overdoped superconductors.
\begin{figure}[b!]
\centerline{\epsfig{figure=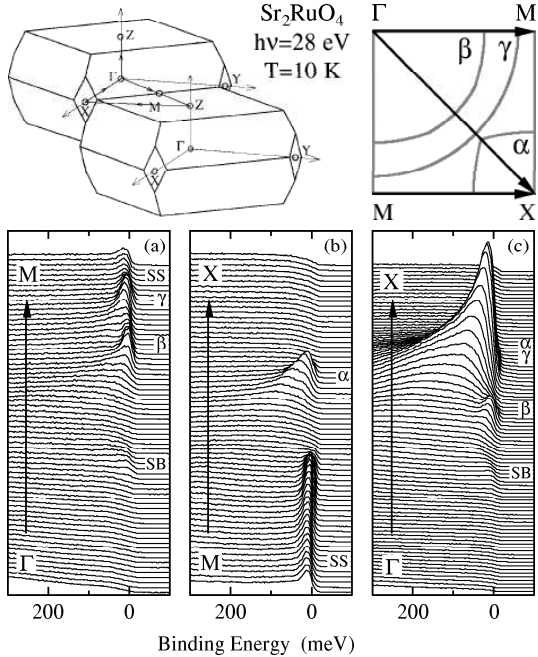,width=0.93\linewidth,clip=}}
\vspace{-.2cm} \caption{ARPES spectra from Sr$_2$RuO$_4$ along the
high symmetry lines $\Gamma$-M, M-X, and $\Gamma$-X, as shown in
the sketch depicting 1/4 of the 2D projected zone. Data from
\cite{andrea_sro}.}\label{Andrea_sroEDC}\end{figure}

\section{State-of-the-art ARPES}

In the early stages of the HTSC field, ARPES proved to be very
successful in measuring the normal state Fermi surface (FS), the
superconducting gap, and the symmetry of the order parameter
\cite{zx}. During the past decade, a great deal of effort has been
invested in further improving this technique which now allows for
energy and momentum resolution of, respectively, a few meV and
$\sim\!1$\% of the typical Brillouin zone (BZ) of HTSCs, thus
ushering in a new era in electron spectroscopy and allowing a very
detailed comparison between theory and experiment. To illustrate
the capability of state-of-the-art ARPES, the novel superconductor
Sr$_2$RuO$_4$ is a particularly good example because of its
complex electronic structure and especially because controversy
has plagued the investigation of its FS topology. In addition,
contrary to the cuprate HTSCs, this material can also be
investigated with other techniques, like de Haas-van Alphen (dHvA)
experiments, thus providing a direct comparison with the ARPES
results.

\begin{figure}[t!]
\centerline{\epsfig{figure=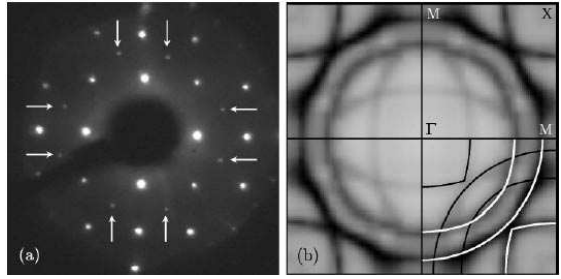,width=0.97\linewidth,clip=}}
\vspace{-.2cm} \caption{(a) LEED pattern obtained with 450 eV
electrons \cite{andrea_sro}. The arrows indicate superlattice
reflections due to $\sqrt2\!\times\!\sqrt2$ surface
reconstruction. (b) $E_F$ intensity map \cite{andrea_sro}. Primary
$\alpha$, $\beta$ and $\gamma$ sheets of FS are marked by white
lines, and replica due to surface reconstruction by black lines.
}\label{Andrea_sroFS}\end{figure}

Whereas dHvA experiments \cite{dhva}, in agreement with LDA
band-structure calculations \cite{lda}, indicate two electron-like
FSs $\beta$ and $\gamma$ centered at the $\Gamma$ point, and a
hole pocket $\alpha$ at the X point (as sketched in 1/4 of the
projected zone in Fig.\,\ref{Andrea_sroEDC}), early ARPES
measurements suggested a different picture: one electron-like FS
at the $\Gamma$ point and two hole pockets at the X point
\cite{lu}. The difference comes from the detection by ARPES of an
intense, weakly dispersive feature at the M point just below
$E_F$, that was interpreted as an extended van Hove singularity
(evHs). Although the evHs was questioned in a later ARPES study
\cite{puchkov}, in which the feature detected at the M point was
suggested to be a surface state (SS), a conclusive picture and a
final agreement between ARPES data, and dHvA and LDA results was
reached only with the `new-generation' of high-resolution
photoemission data.

Fig.\,\ref{Andrea_sroEDC} presents energy distribution curves
(EDCs), along the high-symmetry directions of Sr$_2$RuO$_4$,
recently reported by Damascelli {\it et al.} \cite{andrea_sro}.
Owing to the high momentum and energy resolution (1.5\% of the BZ
and 14 meV), we can now clearly identify several dispersive
features crossing $E_F$ precisely where the $\alpha$, $\beta$, and
$\gamma$ sheets of the FS are expected on the basis of LDA
calculations and dHvA experiments (all detected features in Fig.\
\ref{Andrea_sroEDC} are labeled following their assignment).
Around the M point we can also observe the sharp peak (labeled SS)
that was initially associated with a hole-like sheet of FS
centered at X \cite{lu}.

A Fermi energy intensity map can be obtained by integrating the
EDCs over a narrow energy window about $E_F$ ($\pm10$ meV). As the
spectral function multiplied by the Fermi function reaches its
maximum at $E_F$ when a band crosses the Fermi energy, the FS is
identified by the local maxima of the intensity map. Following
this method, the $\alpha$, $\beta$, and $\gamma$ sheets of FS are
clearly resolved, and are marked by white lines in
Fig.\,\ref{Andrea_sroFS}b. In addition, we find some unexpected
features: weak, yet well defined profiles marked by black lines.
They can be recognized as a replica of the primary FS, and are
related to the weak `shadow bands' (SB) which show dispersion
opposite to the primary peaks along $\Gamma$-M and $\Gamma$-X (see
Fig.\,\ref{Andrea_sroEDC}). The origin of the shadow bands as well
as of the SS, can be identified with the intrinsic instability of
the cleaved surface of Sr$_2$RuO$_4$: inspection with LEED reveals
superlattice reflections corresponding to a
$\sqrt2\!\times\!\sqrt2$ surface reconstruction (Fig.\
\ref{Andrea_sroFS}a), which is responsible for the folding of the
primary electronic structure with respect to the M-M direction
\cite{andrea_sro,plummer}. In light of these findings, the {\it
bulk} FS determined by ARPES is consistent with the LDA and dHvA
results. In addition, ARPES provides essential information on the
detailed shape of the $\alpha$, $\beta$, and $\gamma$ sheets of FS
\cite{andrea_sro}.
\begin{figure}[b!]
\centerline{\epsfig{figure=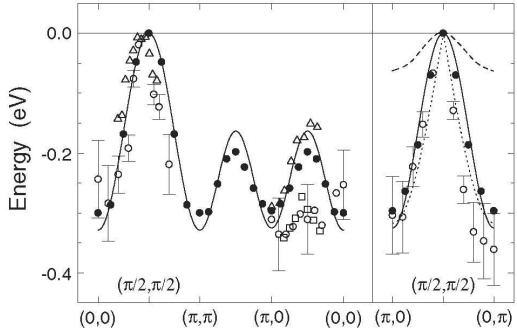,width=0.9\linewidth,clip=}}
\vspace{-.2cm} \caption{Valence band dispersion for SCOC measured
from the top of the band. Experimental data are taken from
\protect{\cite{wells}} (open circles), \protect{\cite{larosa}}
(open triangles) and \protect{\cite{kim}} (open squares). Dashed
line: results from the $t$-$J$ model \protect{\cite{wells}}. Solid
circles: self-consistent Born approximation (SCBA) for the
$t$-$t'$-$t''$-$J$ model ($t$=0.35~eV, $t'$=$-$0.12~eV,
$t''$=0.08~eV and $J$=0.14~eV); solid lines are obtained by
fitting the SCBA data \protect{\cite{tohyama1}}. The dotted line
along the ($\pi$,0)-(0,$\pi$) direction represents the spinon
dispersion given in
\protect{\cite{laughlin}}.}\label{tohyama}\end{figure}

\section{The Mott insulator}

The  $t$-$J$ model, briefly discussed in the introduction, is of
particular relevance to the low-energy features detected in ARPES
on the cuprates. In fact, in ARPES experiments performed on the
insulating parent compounds of the HTSCs, one photo-hole is
injected in the CuO$_2$ plane as a result of the photoemission
process. Therefore, this experiment is the practical realization
of a `single hole' in the AF insulator, and the comparison of
ARPES data and calculations based on the $t$-$J$ model is
particularly meaningful because the single-hole calculation is
free from complications such as charge ordering, which is
difficult for small-cluster calculations to deal with.

Experimental data taken from Ref.\,\cite{wells,larosa,kim} for the
energy dispersion of the quasiparticle (QP) peak in insulating
Sr$_2$CuO$_2$Cl$_2$ (SCOC) are shown in Fig.\,\ref{tohyama} (open
symbols). Note that in the course of the paper we will use terms
like QP in a loose sense for convenience, even though in most
cases FL theory does not apply and well defined QP peaks cannot be
identified in the ARPES spectra. The dispersion along the
(0,0)-($\pi$,$\pi$) direction is characterized by a bandwidth
$W\!\simeq\!0.3$ eV. As pointed out in the first ARPES
investigation on this compound \cite{wells}, this result is in
very good agreement with $t$-$J$ model calculations \cite{liu}
which show that, independent of the value of $t$, the dressing of
the hole moving in the AF background reduces the QP bandwidth from
$8t$ ($\sim\!3$ eV for a free hole) to 2.2$J$ (with
$J\!\simeq\!125$ meV in SCOC, as independently deduced from
neutron scattering studies \cite{greven}). On the other hand, the
$t$-$J$ model also predicts a relatively flat dispersion
\cite{liu} along the ($\pi$,0)-(0,$\pi$) direction (dashed line in
Fig.\,\ref{tohyama}), in contradiction to the more isotropic
dispersion observed in ARPES around ($\pi$/2,$\pi$/2), with
$W\!\simeq\!0.3$ eV independent of the direction. Also the poorly
defined lineshape and the spectral weight suppression observed at
($\pi$,0), which indicate the lack of integrity of the QP at those
momenta, cannot be reproduced within the simple $t$-$J$ model
\cite{kim}.

Better agreement between the experimental dispersion and the
calculations (solid circles and solid line in Fig.\,\ref{tohyama})
is obtained by adding second and third {\it nn} hopping ($t'$ and
$t''$, respectively) to the $t$-$J$ Hamiltonian
\cite{kim,Nazarenko,Kyung1,Xiang,Belinicher,eder,LeeTK,Lema,Leung,Sushkov,tohyama1}.
In fact, as $t'$ and $t''$ describe hopping within the same
magnetic sublattice, they do not alter the AF properties of the
model at half filling; at the same time, they are not strongly
renormalized by the AF correlations but contribute directly to the
coherent motion of the hole and, therefore, have a substantial
impact on the QP dispersion. The inclusion of these terms also
helps in reproducing the suppression, as compared to
($\pi$/2,$\pi$/2), of the QP peak observed in ARPES at ($\pi$,0).
However, nothing can be said about the line shape because the
broadening is artificially introduced in the theory, which is a
major limitation of this kind of approach. Most importantly, it
can be shown that the suppression of the QP peak at ($\pi$,0)
reflects a reduction of AF spin correlations: the additional
hopping possibilities represented by $t'$ and $t''$ induce a
spin-liquid state around the photo-hole with momentum ($\pi$,0)
\cite{tohyama1}. As a consequence, one may expect to find in the
ARPES results some signatures of spin-charge separation
\cite{rvb}. Within this context, it is interesting to note that
the full QP dispersion observed for SCOC can be very well
reproduced also by the spinon dispersion given in \cite{laughlin}
[the dotted line in Fig.\,\ref{tohyama} shows the result along
($\pi$,0)-(0,$\pi$)]. In this case Laughlin argues in favor of the
decay of the photo-hole injected in the 2D\,AF CuO$_2$ plane into
a spinon-holon pair \cite{laughlin}, which is reminiscent of the
flux phase physics
\cite{affleck,wen,chakravarty,maekawa88,fukuyama88,kotliar88}, an
extension of the early resonating valence bond (RVB) conjecture
\cite{rvb}.

The discussion of the ARPES result on insulating SCOC emphasizes a
fundamental problem in the theoretical description of the doped 2D
AF: the Heisenberg model is so strongly perturbed by the addition
of mobile holes that, above a certain doping level, some form of
spin liquid may be a better {\it ansatz} than the long range
ordered N\'{e}el state. This point is centrally important to
high-$T_c$ superconductivity because HTSCs, which are poor
conductors in the normal state, may be better regarded as doped
AFs, whose behavior differs fundamentally from the FL paradigm.
For this reason, the Bardeen-Cooper-Schrieffer (BCS) theory
\cite{bcs} which was developed for Fermi-liquid metals (i.e., weak
electron correlations), and has been so successful in describing
conventional superconductors, does not have the appropriate
foundation for HTSCs. A new approach may therefore be needed, and
a necessary requirement for any theory aiming to capture the
essential physics of high-$T_c$ superconductivity must be the
inclusion of the essential physics of the doped AF: the
competition between AF and Coulomb interactions (which induce
localization), and zero point kinetic energy (which favors
delocalization). Along this direction, the most radical models
seem to be those based on: (i) the RVB state and the related
spin-charge separation picture
\cite{rvb,laughlin,affleck,wen,chakravarty,maekawa88,fukuyama88,kotliar88,ioffe96,anderson_sc,fisher,lee2000},
(ii) stripes
\cite{zaanen,emery,salkola,bianconi96,machida99,tohyama_str,scalapino00,markiewicz00,fleck00,castro00,zacher1,han},
and (iii) quantum criticality
\cite{chakravarty89,sachdev92,sokol93,emery93,castellani95,varma97}.
Independent of their details, these different theoretical
approaches have one important common denominator:
superconductivity is not caused by the pairing of two QPs, as in
the BCS case, rather it is the process in which the QP itself
forms. Furthermore, in the first two cases the driving mechanism
for the superconducting phase transition is identified with the
gain in kinetic energy, contrary to the standard theories of
solids where any phase transition into a long-range ordered state
is driven by the gain in potential energy. In the stripe or RVB
models the hopping of pairs of holes perturbs the AF spin
background less than individual holes. However, it is only when
charge fluctuations become phase coherent that the frustration of
the kinetic energy is released, and superconductivity sets in.

\subsection{Remnant FS and $d$-wave-like dispersion}

\begin{figure}[t!]
\centerline{\epsfig{figure=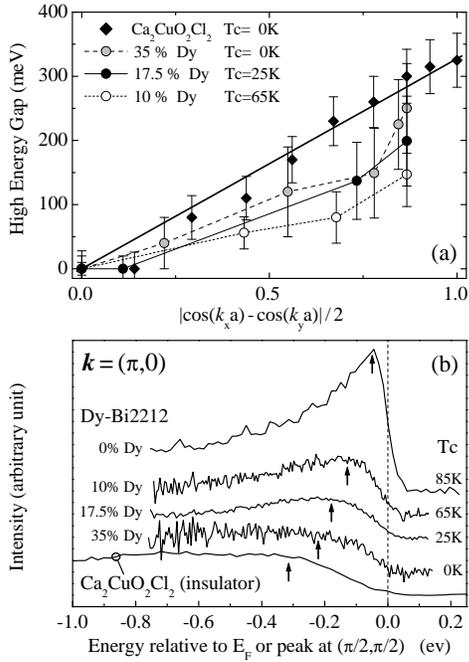,width=0.82\linewidth,clip=}}
\vspace{-.2cm} \caption{(a) High energy pseudogap versus $|\!\cos
k_x\!-\! \cos k_y|/2$ for CCOC, and Dy-Bi2212. (b) Doping
dependence of the ($\pi$,0) ARPES spectra. Data taken from
\cite{filip}.}\label{Filip_DW}\end{figure}

We mentioned above  that both the relatively isotropic dispersion
at ($\pi$/2,$\pi$/2), and the suppression of QP weight at
($\pi$,0), observed from ARPES on SCOC and Ca$_2$CuO$_2$Cl$_2$
(CCOC, similar in many respects to SCOC \cite{filip}), cannot be
explained with the {\it nn} hopping $t$-$J$ model. Better
agreement with the experiment is obtained by including longer
range hopping terms in the model. In this way, it is possible to
also reproduce the doping dependence of the QP band structure and,
in particular, of the ($\pi$,0) ARPES spectra \cite{eder}. These
are shown for optimally doped Bi$_2$Sr$_2$CaCu$_2$O$_8$ (Bi2212),
Dy-Bi2212 and CCOC \cite{filip} in Fig.\,\ref{Filip_DW}b. Note
that for CCOC the zero in energy does not correspond to $E_F$ but
to the peak position at ($\pi$/2,$\pi$/2) which, because of the
Mott gap, is located $\sim\!700$ meV below the chemical potential
and corresponds to the top of the valence band.
In underdoped samples the QP peak at ($\pi$,0) loses coherence and
shifts to higher BE, while at ($\pi$/2,$\pi$/2) spectral weight
still reaches the Fermi level. The lack of a FS crossing along the
($\pi$,0)-($\pi$,$\pi$) cut in the normal state of the underdoped
regime indicates the opening of a pseudogap
\cite{filip,loeser,marshall96,dingN96}, which appears to be
characterized by the same $d$-wave symmetry found for the
superconducting gap \cite{dwave}. Because the maximum size of the
pseudogap is found at the $E_F$ crossing along the
($\pi$,0)-($\pi$,$\pi$) direction, and the dispersion around
($\pi$,0) is very weak, the doping dependence of the pseudogap
magnitude can be deduced from the energy position of the QP peak
at ($\pi$,0), as shown in Fig.\,\ref{Filip_DW}b.

However, the  $t$-$J$ model, even in its more extended form,
cannot completely account for the strong momentum dependence of
the ARPES spectra from the undoped insulator \cite{eskes}. In
particular, even though a drop of intensity of the lowest energy
peak along (0,0)-($\pi$,$\pi$), after the AF BZ boundary, is
predicted by the $t$-$J$ model \cite{eskes,bulut}, it is not as
sharp as experimentally observed in SCOC and CCOC
\cite{wells,filip}. This limitation of the $t$-$J$ model comes
from having projected out the doubly occupied states originally
contained in the Hubbard model: whereas the momentum occupation
number $n(k)$ is a strongly varying function of $k$ in the
intermediate-$U$ Hubbard model at half filling, it is trivially
equal to 1/2 in the $t$-$J$ model which, therefore, cannot
describe the anomalous distribution of spectral weight in the
single particle spectral function. This effect is accounted for by
the complete large U-limit of the Hubbard model (i.e., without
neglecting certain terms in the large $U$ perturbation theory), as
shown by Eskes and Eder \cite{eskes}. For $U\!\rightarrow\!\infty$
the spectra of large-U-Hubbard and $t$-$J$ model will coincide,
and $n(k)\!\rightarrow\!1/2$; however, the convergence in $U$ is
very slow. On the basis of finite size cluster calculations within
the full strong coupling model, Eskes and Eder reproduced the
sharp drop of intensity observed in the ARPES spectra from SCOC at
the AF BZ boundary (referred to, by the same authors, as a
`pseudo-FS' \cite{eskes}).

\begin{figure}[t!]
\centerline{\epsfig{figure=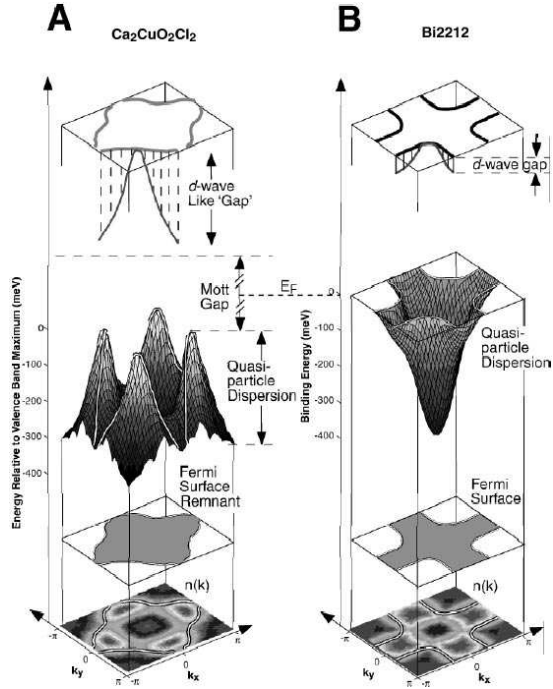,width=0.95\linewidth,clip=}}
\vspace{-.2cm} \caption{FS and RFS (bottom) defined by the
analysis of $n(k)$ for overdoped Bi2212, and insulating CCOC. Note
that while the FS in Bi2212 is the isoenergetic contour located at
$E_F$, the RFS is away from $E_F$ (because of the presence of the
Mott gap), and a large $d$-wave-like dispersion (300 meV) is found
along its contour. The latter defines a $d$-wave `gap' for the
insulator (top left), similar to the $d$-wave pseudo gap observed
in the underdoped regime (top right), and due to strong
correlations which deform the isoenergetic FS of the overdoped
metal (from \cite{filip}).}\label{Filip_cartoon}\end{figure}

A detailed experimental characterization of the $k$-dependence of
the ARPES spectral weight for the undoped insulator has been
presented by Ronning {\it et al.} \cite{filip}, on the basis of
$n(k)$ mapping obtained by integrating the EDCs from CCOC over an
energy window larger than the bandwidth. From the location of the
steepest drops in $n(k)$, which successfully gives the FS for
overdoped Bi2212, they could define a `remnant-FS' (RFS) for CCOC
(Fig.\,\ref{Filip_cartoon}a and\,\ref{Filip_cartoon}b, bottom),
which closely follows the AF BZ boundary. Note that matrix element
effects also influence the $k$-dependence of the intensity and
alter the profile of the RFS \cite{haffner}; however the sharp
drop at the AF BZ boundary observed at different photon energies,
in the experiment, and in the numerical results appears to be a
robust feature, despite minor uncertainties related to the photon
energy dependence of the photoionization cross section
\cite{filipcond,fink}.

It is important to realize that the RFS is not a real FS (the
system has a Mott gap) but identifies the same locus of rapid
intensity drop referred to as a pseudo-FS by Eskes and Eder
\cite{eskes}. In addition, it does not even correspond to an
isoenergetic contour in the QP dispersion
(Fig.\,\ref{Filip_cartoon}), similarly to the FS determined in the
underdoped regime. The relevance of this approach is that, once a
RFS has been determined, it is also possible to identify a `gap'
along its contour (in addition to the Mott gap), and try to
compare it to the high energy pseudogap of the underdoped systems.
As reported by Ronning {\it et al.} \cite{filip}, the `high
energy' pseudogap (given by the position of the broad peak
indicated by arrows in Fig.\,\ref{Filip_DW}b) shows $d$-wave
symmetry not only in the underdoped systems but also in the
undoped insulator. This is shown in Fig.\,\ref{Filip_DW}a where
the dispersion of the high energy pseudogap along the FS (RFS for
CCOC) is plotted against the $d$-wave functional form (a fit for
CCOC is also shown, whereas the other lines are only guides to the
eye). Although their sizes vary, the superconducting gap, the
pseudogap of the underdoped system, and the gap of the insulator
have the same nontrivial $d$-wave form, suggesting a common origin
\cite{zacher}. This is consistent with the idea of one underlying
symmetry principle [i.e., SO(5)] that unifies the AF insulating
state and the $d$-wave superconducting state \cite{zhang1}.

\section{Evolution of the electronic structure}

\begin{figure}[t!]
\centerline{\epsfig{figure=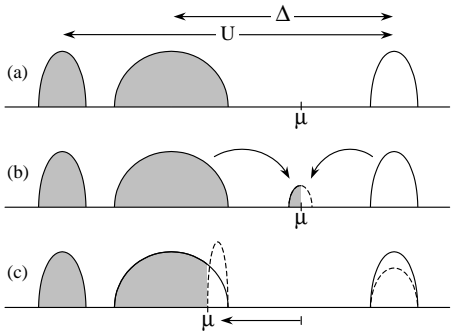,width=0.8\linewidth,clip=}}
\vspace{-.2cm}\caption{Doping of a charge-transfer insulator (a):
$\mu$ is pinned inside the charge-transfer gap and states move
towards the chemical potential (b); alternatively, $\mu$ shifts to
the top of the valence band and spectral weight is transferred
because of correlations (c). Figure taken from
\cite{sawa}.}\label{mott}\end{figure}

The work on the undoped insulator provides a starting point to
understand the doping evolution of the electronic structure of the
cuprates. Upon doping the system, AF correlations are reduced and
a metallic state appears. Eventually (i.e., in the optimum and
overdoped regime), the AF state is destroyed and a large LDA-like
FS appears \cite{zx}, with a volume which scales as ($1\!-\!x$),
counting electrons  ($x$ is the concentration of doped holes for
p-type HTSCs), as expected within the FL approach. In this
context, the first important question to answer concerns the way
the low energy states emerge in the underdoped regime. For
$x\!\ll\!1$, two alternative scenarios have been proposed (see
Fig.\,\ref{mott} for hole doping): (i) the chemical potential
$\mu$ is pinned inside the charge-transfer gap $\Delta$, as
`in-gap states' are created \cite{allen} (Fig.\,\ref{mott}b); (ii)
the chemical potential moves downwards into the top of the valence
band and states are transferred from the UHB to the LHB because of
correlations \cite{sawa} (Fig.\,\ref{mott}c). Another relevant
question is how the low-lying states evolve from underdoped to
overdoped regime, where FL behavior seems to recover \cite{zx}. To
better organize the discussion, let us layout some relevant
theoretical models. They can be classified as: (i) those that
preserve the underlying crystalline symmetry; (ii) those that
break this symmetry. Note that also those scenarios based on a
dynamical breaking of symmetry should be taken into account
because ARPES is sensitive to the latter, due to the relatively
high excitation energy.

\begin{figure}[b!]
\centerline{\epsfig{figure=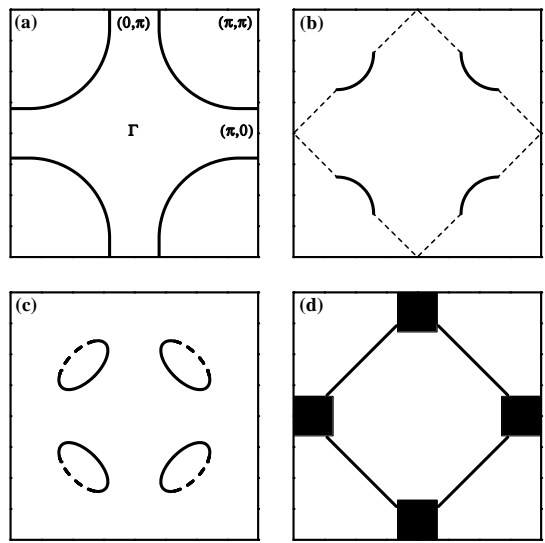,width=0.95\linewidth,clip=}}
\vspace{-.2cm}\caption{FS obtained for the CuO$_2$ plane from (a)
LDA \cite{zx}, (b) truncation of a 2D FS due to {\it umklapp}
scattering \cite{rice}, (c) RVB/flux-phase approach \cite{wen},
and (d) stripe model for vertical and horizontal domains of
disordered stripes \cite{salkola}.}\label{models}\end{figure}

The first models to be mentioned among (i) are the FL and band
structure perspectives \cite{pines,pickett89}, which sever the
connection to the undoped AF insulator by assuming that the
screening in the doped metal is strong enough for the FL formalism
to recover; in this case a well defined FS is expected
(Fig.\,\ref{models}a), with a volume proportional to ($1\!-\!x$).
An alternative scenario considers the breakdown of FL theory due
to {\it umklapp} scattering \cite{rice}. As a consequence, in the
underdoped region of the phase diagram, the FS is truncated near
the saddle points at ($\pi$,0) and (0,$\pi$) because of the
opening of spin and charge gaps. This results in four disconnected
arcs of FS centered at ($\pm\pi$/2,$\pm\pi$/2), as shown in
Fig.\,\ref{models}b. In agreement with a generalized form of
Luttinger's theorem, the area defined by the four arcs and by the
{\it umklapp} gapped FS (dashed lines in Fig.\,\ref{models}b)
encloses the full electron density.

Among the broken-symmetry models, we find the RVB/flux-phase
approach
\cite{affleck,wen,chakravarty,maekawa88,fukuyama88,kotliar88} that
predicts a FS given by four hole-pockets close to
($\pm\pi$/2,$\pm\pi$/2), as in Fig.\,\ref{models}c, which
continuously evolve into a large FS upon increasing the hole
concentration. This is in a very similar spirit to that of the
spin-density wave picture which also assumes a dynamical breaking
of symmetry \cite{kampf}. Another model belonging to (ii) is the
stripe picture \cite{salkola}, which yields a momentum-space
distribution of low-lying excitations. These are represented by
the black patches in Fig.\,\ref{models}d, where the results
obtained for an equal number of vertical and horizontal domains of
disordered stripes are qualitatively sketched (in this case the
physics, together with the superposition of domains, conspires to
give the appearance of a large LDA-like FS).
\begin{figure*}[t!]
\centerline{\epsfig{figure=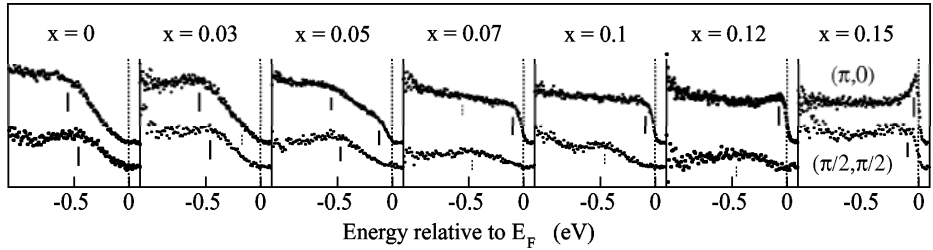,width=0.77\linewidth,clip=}}
\vspace{-.3cm}\caption{Doping dependence of the ARPES spectra for
LSCO at ($\pi$,0) and ($\pi/2$,$\pi/2$). EDCs were normalized to
the integrated intensity of the valence bands ($E\!>\!-0.9$ eV),
and for ($\pi/2$,$\pi/2$) were multiplied by a factor of two. Data
taken from \cite{ino_nodal}.}\label{Akihiro_Dop}\end{figure*}

There is actually another meaningful way to differentiate the four
models discussed above: those depicted in
Fig.\,\ref{models}a,\,\ref{models}b, and\,\ref{models}c assume
that the system is spatially uniform (as far as
Fig.\,\ref{models}b is concerned, one could talk about phase
separation between insulating spin-liquid and metallic regions,
but only in momentum space \cite{rice}). To the contrary, the
model in Fig.\,\ref{models}d assumes that the system is spatially
non-uniform: the formation of stripes is defined as the
segregation of charge carriers into 1D domain walls which separate
AF spin domains in antiphase with each other, and here, in
particular, disordered stripes are considered
\cite{salkola,tranquada}.

Each of the above discussed pictures captures some aspects of the
experimental reality. In the course of the paper, we will try to
compare ARPES data from various systems with the results of these
models, aiming to identify the scenario that has the best overlap
with experimental observations. This will also help us to answer
the question whether different materials would favor different
scenarios, and to address the relevance of degrees of freedom
other than the electronic ones (e.g., lattice degrees of freedom
in the case of the stripe instability).

\subsection{La$_{2-x}$Sr$_x$CuO$_4$}

In order to study the doping evolution of the low-energy
electronic properties over the full doping range, in particular in
the vicinity of the metal-insulator transition (MIT), the most
suitable system is LSCO. The hole concentration in the CuO$_2$
plane can be controlled and determined by the Sr content $x$, from
the undoped insulator ($x\!=\!0$) to the heavily overdoped metal
($x\!\sim\!0.35$). In addition, LSCO has a simple crystal
structure with a single CuO$_2$ layer, and no complications due to
superstructure and shadow bands, as in the case of Bi2212
\cite{zx}. Another interesting aspect is the suppression of $T_c$
at $x\!=\!1/8$ which, together with the incommensurate AF
long-range order observed in inelastic neutron scattering
\cite{suzuki}, has been discussed as evidence for fluctuating
stripes in LSCO (similar AF order accompanied by charge ordering
has been interpreted as a realization of `static stripes' in
La$_{1.48}$Nd$_{0.4}$Sr$_{0.12}$CuO$_4$ \cite{tranquada}).

Let us start from the low doping region, near the MIT boundary.
Fig.\,\ref{Akihiro_Dop} presents the ARPES spectra at ($\pi$,0)
and ($\pi$/2,$\pi$/2) as a function of doping, reported by Ino
{\it et al.} \cite{ino_nodal}. The data was recorded under
identical experimental geometry so that the photoionization matrix
elements are the same. For the insulating samples
($x\!\leq\!0.03$), the data is characterized by a high binding
energy (BE) feature [$\sim\!0.5$ eV at ($\pi/2$,$\pi/2$), and
$\sim\!0.7$ eV at ($\pi$,0)], and is consistent with what we have
discussed in the last section for insulating SCOC
\cite{wells,kim}, albeit the features are now broader. The
remarkable result is that for $x\!=\!0.05$ two features can be
identified in the EDC at ($\pi$,0): in addition to the high BE
one, reminiscent of the ZR singlet band of the AF insulator, a
second shoulder is observable close to $E_F$. Upon further doping
the system with holes, a systematic transfer of spectral weight
from the high-BE to the low-BE feature takes place, and a
well-defined QP peak develops near optimal doping. On the other
hand, the results obtained at ($\pi/2$,$\pi/2$) are very
different: first of all, the data shows an overall suppression of
weight as compared to ($\pi$,0) [the EDCs plotted in
Fig.\,\ref{Akihiro_Dop} for ($\pi/2$,$\pi/2$) have been multiplied
by a factor of 2]; second, in the nodal region [i.e., along
(0,0)-($\pi$,$\pi$)], a QP peak is observable only for
$x\!\geq\!0.15$ \cite{ino_nodal}. As we will discuss later, with
different experimental geometries more spectral weight is detected
in the nodal region, but the overall trend of the doping
dependence of the electronic structure is robust.

\begin{figure}[b!]
\centerline{\epsfig{figure=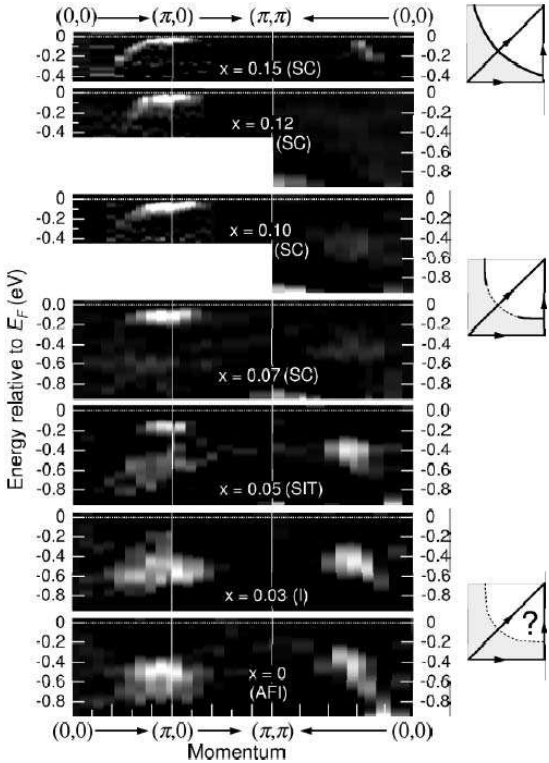,width=0.95\linewidth,clip=}}
\vspace{-.2cm}\caption{Second derivatives of the ARPES spectra
from LSCO over a broad doping range. Data taken from
\cite{ino_nodal}.}\label{Akihiro_Der}\end{figure}

The overall dispersion of the spectral features seen in LSCO and
their doping dependence is summarized by the plot of the second
derivative (taken with respect to the energy) of the ARPES spectra
presented in Fig.\,\ref{Akihiro_Der}. Upon increasing the doping
level, we can clearly observe the building of near-$E_F$ weight
first at ($\pi$,0), and then at ($\pi$/2,$\pi$/2). Furthermore,
the second derivative emphasizes the presence of the high-BE
feature in the heavily underdoped samples (coexisting with the low
BE one, at least for $x\!=\!0.05$), which has a $\sim\!200$ meV
lower BE at ($\pi$/2,$\pi$/2) than at ($\pi$,0), in agreement with
what is observed on the undoped insulator SCOC \cite{wells}.
The ARPES results from LSCO, in particular the presence of two
electronic components and the fact that the low BE feature emerges
first at ($\pi$,0) and then at ($\pi$/2,$\pi$/2), suggest that the
effects of doping on the electronic structure of the correlated
insulator cannot be accounted for by a simple shift of the Fermi
level in a rigid band model \cite{ino_nodal}. In fact, in the
latter case the lowest excitations would first appear at
($\pi$/2,$\pi$/2). This argument is in agreement with another
observation by Ino {\it et al.} \cite{ino_mu}: from the analysis
of direct and inverse angle-integrated photoemission spectra, they
concluded that the chemical potential $\mu$ is pinned inside the
charge-transfer gap for $x\!\leq\!0.12$, and starts shifting
downwards only for $x\!\geq\!0.12$. All the above results seem to
indicate that in-gap states are created upon doping the insulator
(Fig.\,\ref{mott}b).

Ino {\it et al.} argued that the ARPES results from LSCO may be
understood within the stripe picture \cite{ino_nodal}. This would
explain the pinning of the chemical potential for $x\!\leq\!1/8$
as a consequence of the segregation of doped holes into metallic
domain walls, which corresponds to the appearance of in-gap
states. Furthermore, the suppression of nodal intensity at $E_F$
would be a direct consequence of the vertical and horizontal
orientation of the metallic stripes (domains are expected for this
microscopic phase separation) \cite{tohyama_str}. Here the
conjecture is that charge fluctuations would be suppressed along
directions crossing the stripes, and is supported by finite-size
cluster calculations \cite{tohyama_str}. The sudden increase of
$E_F$ weight for $x\!\geq\!1/8$ in the nodal region may indicate
that above this doping level the holes overflow from the saturated
stripes.

\begin{figure}[t!]
\centerline{\epsfig{figure=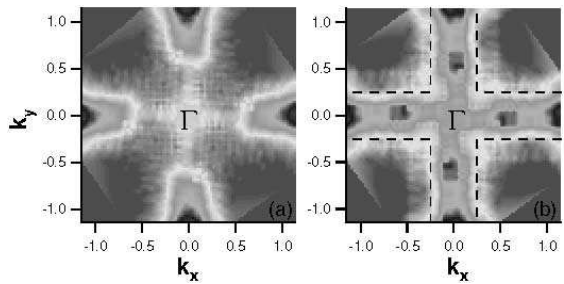,width=0.975\linewidth,clip=}}
\vspace{-.2cm} \caption{Intensity maps obtained by integrating the
EDCs over 100 meV (a), and 500 meV (b) below $E_F$. White lines in
(b) enclose the high intensity region. Data from
\cite{zhou}.}\label{Xingjiang_FS}\end{figure}

Concerning the relevancy of the stripe scenario to the ARPES data
from HTSCs, more insights could come from the investigation of
Nd-LSCO, a model compound for which the evidence of spin and
charge stripe-ordering is the strongest \cite{tranquada}. High
momentum resolution ARPES data on
La$_{1.28}$Nd$_{0.6}$Sr$_{0.12}$CuO$_4$ were recently reported by
Zhou {\it et al.} \cite{zhou}, and are shown in
Fig.\,\ref{Xingjiang_FS}, where spectral weight maps obtained by
integrating the EDCs over 100 and 500 meV energy windows below
$E_F$ are presented (Fig.\,\ref{Xingjiang_FS}a
and\,\ref{Xingjiang_FS}b, respectively). The data was symmetrized
in accordance with the four fold symmetry of the BZ [note that the
symmetry with respect to the (0,0)-($\pi$,$\pi$) line is directly
observed in the raw data taken, in the same experimental geometry,
over a whole quadrant of the BZ]. The low-energy spectral weight
is mostly concentrated in a narrow region along the
(0,0)-($\pi$,0) direction, which is confined between lines
crossing the axes at $\pm\,\pi/4$ \cite{zhou}.

\begin{figure}[t!]
\centerline{\epsfig{figure=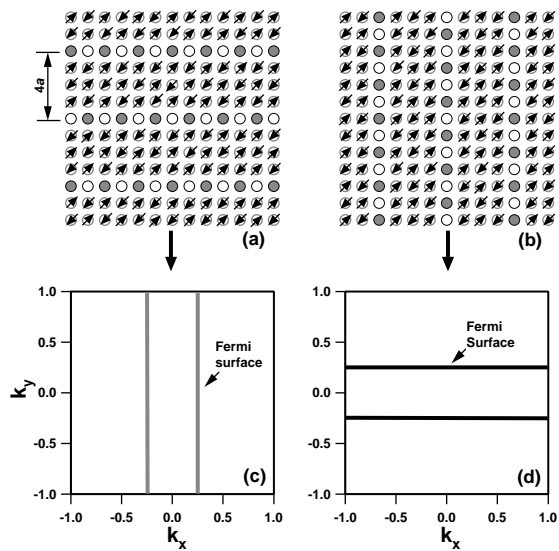,width=0.975\linewidth,clip=}}
\vspace{-.2cm} \caption{Horizontal (a) and vertical (b) static
stripes, and their corresponding FS expected to be defined by the
lines $|k_x|\!=\!\pi/4$ (c), and $|k_y|\!=\!\pi/4$ (d),
respectively (from \cite{zhou}).}\label{Xingjiang_1D}\end{figure}

The high intensity area of the BZ in Fig.\,\ref{Xingjiang_FS}b,
which is suggestive of almost perfectly nested 1D FS segments, is
consistent with the 1/8 doping level of the system. Zhou {\it et
al.} \cite{zhou} interpret these results as a signature of a 1D
electronic structure related to the presence of static stripes. As
indicated by neutron and x-ray experiments \cite{tranquada}, at
1/8 doping the 1/4-filled charge stripes are separated by AF
domains with a periodicity of 4$a$ (Fig.\,\ref{Xingjiang_1D}),
where $a$ is the lattice parameter. This picture is also
consistent with various theoretical calculations
\cite{machida99,markiewicz00,castro00}. In particular, the
preponderance of low lying excitations at ($\pi$,0), which is
observable in Fig.\,\ref{Xingjiang_FS}a, is consistent with
calculations for disordered stripes like, e.g., those summarized
in Fig.\,\ref{models}d \cite{salkola,zacher1}. Concerning the
macroscopic orientation of the stripes, two orthogonal domains are
expected, as shown in Fig.\,\ref{Xingjiang_1D}a and
\ref{Xingjiang_1D}b. For each domain the FS consists of straight
lines perpendicular to the direction of the charge stripe itself,
and intersecting the axes of the BZ at $\pm\,\pi/4$
(Fig.\,\ref{Xingjiang_1D}c and \ref{Xingjiang_1D}d). The intensity
distribution observed in Fig.\,\ref{Xingjiang_FS}b, would then
result from the superposition of two perpendicular FSs reflecting
the presence of orthogonal domains.

The above interpretation of the ARPES data on Nd-LSCO would also
provide a possible explanation for the origin of the two
components seen in the ARPES spectra of LSCO near the MIT boundary
(Fig.\,\ref{Akihiro_Dop}). In the static picture discussed above,
the signal from the AF insulating regions would be pushed to high
BE because of the Mott gap, whereas the charge stripes would be
responsible for the component near $E_F$. In this sense, the
stripe interpretation is rather appealing. On the other hand, the
picture discussed in \cite{zhou} is based on the assumption of an
extreme charge disproportionation which is usually not found in
charge ordering transitions, even in lower symmetry systems such
as ladders \cite{cristian}. Nevertheless, the qualitative picture
presented in Fig.\,\ref{Xingjiang_1D} may capture the relevant
physics even for a less extreme case of charge disproportionation.

There are also some results which cannot be satisfactorily
explained within the framework of static stripes. For example, in
both LSCO (Fig.\,\ref{Akihiro_Der}) and Nd-LSCO \cite{zhou}, the
QP band along the (0,0)-($\pi$,0) direction is characterized by a
considerably fast dispersion, contrary to what is expected for an
ideal 1D system which typically does not exhibit any dispersion
perpendicularly to its main axis. Furthermore, matrix element
effects have to be cautiously considered when interpreting ARPES
data, especially in dealing with the integrated spectral weight.
In fact, although the integration of the EDCs over a large energy
window gives an estimate for the momentum-dependent occupation
number $n(k)\!=\!\int\!A(k,\omega)f(\omega)d\omega$, the latter
quantity is weighted by the photoionization cross section and thus
may contain extrinsic artifacts \cite{hufner}.
\begin{figure}[t!]
\centerline{\epsfig{figure=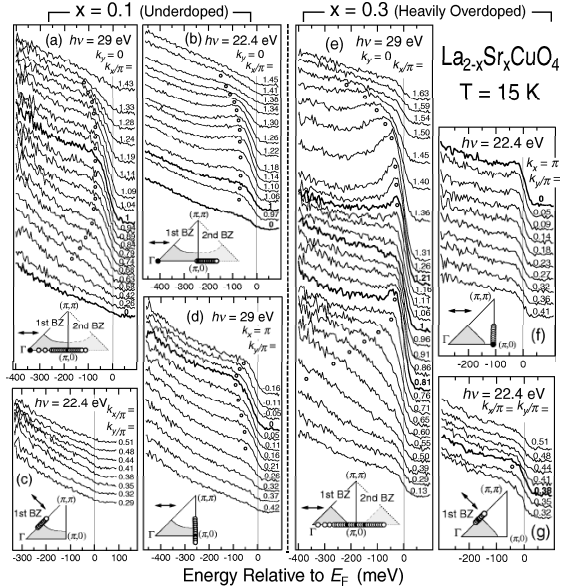,width=0.975\linewidth,clip=}}
\vspace{-.2cm}\caption{ARPES spectra of underdoped ($x\!=\!0.1$)
and heavily overdoped ($x\!=\!0.3$) LSCO, from \cite{ino_dop}.
Insets show measured $k$-space points and incident light
polarization.}\label{Akihiro_EDC}\end{figure}
In order to gain more insight into this issue, in particular in
relation to the straight segments of FS observed in Nd-LSCO and to
the suppression of the nodal state, Zhou {\it et al.} \cite{zhou1}
extended the measurements to the second zone, and varied
polarization and orientation of the incoming electric field to
enhance the spectral intensity in the ($\pi$/2,$\pi$/2) region. As
a result, the presence of the nested segments of FS near ($\pi$,0)
and (0,$\pi$) was confirmed. On the other hand, appreciable
spectral weight at the Fermi level was found in the nodal region
\cite{zhou1}, which appears to become more intense upon increasing
the Sr concentration in both Nd-LSCO and LSCO, and is stronger in
Nd-free LSCO for a given Sr content \cite{zhou1}. A possible way
of understanding these results within the stripe context, as
suggested by Zhou {\it et al.} \cite{zhou1}, is that the nodal
state and the dispersion along (0,0)-($\pi$,0) may arise from
disorder or fluctuation of the stripes, where the holes leak into
the AF regions \cite{salkola}. As a matter of fact, the detected
FS  \cite{zhou1}, which is composed by straight patches at
($\pi$,0) and (0,$\pi$) connected by `nodal segments', closely
resembles the one depicted in Fig.\,\ref{models}d. Alternatively,
this experimental FS may result from the coexistence of
site-centered and bond-centered stripes \cite{zacher1}. Both of
these scenarios suggest that the charge disproportionation is not
extreme, as also indicated by dynamical mean-field calculations
\cite{fleck00}.

\begin{figure}[b!]
\centerline{\epsfig{figure=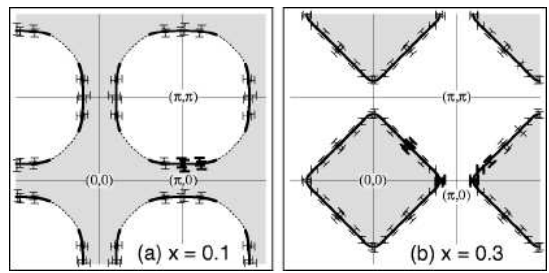,width=0.95\linewidth,clip=}}
\vspace{-.2cm} \caption{FS of LSCO for $x\!=\!0.1$ (a) and
$x\!=\!0.3$ (b), taken from \cite{ino_dop}. Thick (thin) error
bars denote FS crossings observed (folded by symmetry). For
$x\!=\!0.1$, FS crossings correspond to minimum-gap loci; as no
dispersive feature is observed at ($\pi$/2,$\pi$/2) near $E_F$
(see EDCs in Fig.\,\ref{Akihiro_EDC}c), the dotted curve is drawn
so that the FS area is $\sim\!90\%$ of the BZ area (in the respect
of the Luttinger theorem).}\label{Akihiro_FS}\end{figure}

Lastly, it was noted that this composite FS has the appearance of
a large LDA-like FS \cite{zhou1}, like the one depicted in
Fig.\,\ref{models}a. Because of disorder and the fact that the
charge disproportionation is less than in the idealized stripe
model of Fig.\,\ref{Xingjiang_1D}, one may still talk about the FS
as the locus of low-lying excitations in $k$-space. This is
particularly true for highly doped cases as the stripe effect
should disappear. This existence of a FS in a striped system is
found to be true in both cluster perturbation theory
\cite{zacher1} and dynamical mean-filed type of calculations
\cite{salkola}. It is then meaningful to discuss the doping
evolution of this LDA-like FS. ARPES spectra for underdoped
($x\!=\!0.1$) and overdoped ($x\!=\!0.3$) LSCO, reported by Ino
{\it et al.} \cite{ino_dop}, are shown in Fig.\,\ref{Akihiro_EDC}.
Note that the spectral features tend to be broad, which may be
related to charge inhomogeneity. Nevertheless, some Fermi
crossings are still observable in part of the BZ, especially in
overdoped samples. For $x\!=\!0.1$, along the direction
(0,0)-($\pi$,0)-(2$\pi$,0) at 29 eV photon energy
(Fig.\,\ref{Akihiro_EDC}a), a broad QP peak emerges from the
background, disperses towards $E_F$ without crossing it, and then
pulls back in the second BZ. Similar results are obtained at 22.4
eV (Fig.\,\ref{Akihiro_EDC}b), the only difference being a
decrease of intensity in the first BZ due to matrix element
effects specific to this photon energy \cite{ino_dop}. Along
($\pi$,0)-($\pi$,$\pi$) the QP peak (Fig.\,\ref{Akihiro_EDC}d),
with maximum binding energy (BE) at ($\pi$,0), disperses almost up
to $E_F$, loses intensity, and disappears. The leading-edge
midpoint never reaches $E_F$ because of the superconducting gap
($\sim\!8$ meV) opening along the FS at this temperature. In this
case, the underlying FS is identified by the locus of the minimum
gap \cite{campuzano}, located here at ($\pi$,0.2$\pi$). Along the
nodal direction no clear peak can be identified
(Fig.\,\ref{Akihiro_EDC}c), as discussed above. However, having
detected a band below $E_F$ at ($\pi$,0), the authors conclude
that for $x\!=\!0.1$ the FS of LSCO is hole-like in character and
centered at ($\pi$,$\pi$), as shown in Fig.\,\ref{Akihiro_FS}a
\cite{ino_dop}.

By comparing the EDCs from heavily overdoped and underdoped LSCO
(Fig.\,\ref{Akihiro_EDC}e and\,\ref{Akihiro_EDC}a, respectively),
we see a striking difference: for $x\!=\!0.3$ the QP peak present
along this cut has almost disappeared at ($\pi$,0). The decrease
of intensity, together with a leading-edge midpoint now located
above $E_F$, provides evidence for the QP peak crossing $E_F$ just
before ($\pi$,0). The FS thus determined for heavily overdoped
LSCO (Fig.\,\ref{Akihiro_FS}b) is electron-like in character and
is centered at (0,0). Furthermore, careful investigations by Ino
{\it et al.} \cite{ino_dop1} and Zhou {\it et al.} \cite{zhou1}
show that the FS changes from hole-like to electron-like for
$x\simeq0.15$\,-\,0.2.

In summary, what has emerged from the study of the LSCO system is
a very complex and intriguing picture, characterized by some
contrasting aspects: neither a simple stripe model, nor any of
other models proposed in Fig.\,\ref{models} can provide a
satisfactory explanation for the complete body of available data.
As we have discussed, the stripe picture, when
disorder/fluctuations and more realistic charge disproportion are
considered, has the advantage of qualitatively explaining the data
over the entire doping range, including the presence of two
electronic components, the straight FS segments, and the lack of a
chemical potential shift in the very underdoped regime. On the
other hand, on a more quantitative level, there are still many
open questions. On the experimental side, two issues should be
carefully considered. The first one is the role of matrix element
effects. In fact, {\it ab initio} calculations of matrix elements
are still unavailable for LSCO, and the tight binding fits do not
reproduce the results to a satisfactory degree. Therefore, the
most robust information at present may come from the analysis of
the systematic changes observed in data recorded under identical
experimental conditions (e.g, as the doping dependence studies
discussed in this section). Second, the system may consists of
metallic stripes aligned along both [1,0] and [1,1], and
characterized by smaller and larger charge disproportionation,
respectively. If, on the one hand, this scenario might explain the
photoemission data, on the other hand the coexistence of two
phases is observed only near 5\% doping. On the theoretical side,
it is unclear how the quasi-1D electronic structure of the stripe
phase can be smoothly connected to the 2D electronic behavior of
the overdoped regime. Although more effort has to be invested in
the study of the electronic properties of charge-ordered systems
and, in particular, in investigating the role of the
electron-lattice interaction, encouraging numerical studies were
recently reported \cite{fleck00,zacher1}, which suggest that the
spectral properties of this materials show a non trivial
superposition of 2D\,AF and 1D metallic behavior. In particular,
the dispersion observed in ARPES perpendicular to the stripe
direction would stem from the AF domains \cite{fleck00,zacher1}.

\subsection{Bi$_2$Sr$_2$CaCu$_2$O$_{8+\delta}$}

In proceeding with the comparative study of the cuprates, let us
now turn our attention to Bi2212 which is the HTSC system most
intensively investigated by ARPES (thanks to the presence of a
natural cleavage plane between the BiO layers). Due to sample
quality issues, most of the Bi2212 experiments were carried out
near optimal doping, and there is almost no information on the
electronic structure near the MIT boundary. Here we concentrate on
cases with a doping of 10\% or higher. Therefore, we cannot answer
the question whether the two-component electronic structure
observed in the LSCO system with a doping of 5-7\% is also present
in Bi2212 cases. It is still an open question as to how the
metallic state emerges in this system.

Bi2212 can also be considered as the most debated HTSC system as
far as ARPES is concerned, because of the complexity of the
electronic structure near ($\pi$,0). These complications arise
from the detection of additional features besides those related to
the primary electronic structure: `shadow bands' (possibly
reflecting AF correlations \cite{aebi} or the presence of two
formula units per unit cell \cite{zx}), and `{\it umklapp} bands'
(due to the diffraction of the photoelectrons from the
superstructure present in the BiO layers \cite{ding96}). As a
result, around ($\pi$,0) two main bands, two shadow bands, and
four {\it umklapp} bands cross the Fermi level. One additional
problem with Bi2212 is that there are no reliable band
calculations: all theoretical results predicted a BiO FS that has
never been observed \cite{zx}. In the following, we will discuss
the current understanding of these and other more recent issues of
the investigation of Bi2212 and its electronic properties.

\subsubsection{Fermi surface}

After an initial debate, a consensus has been reached concerning
the absence of the BiO pocket, which was predicted for Bi2212 by
band structure calculations. Furthermore, the general consensus is
in favor of a hole-like FS centered at ($\pi$,$\pi$), with a
volume consistent with the electron density in accordance with the
Luttinger theorem \cite{zx,aebi,ding96}. In contrast to an earlier
study that suggested the presence of an electron-like FS due to
bilayer splitting \cite{dessau93}, it was argued that there is no
conclusive evidence for this effect \cite{ding96}. For a period of
time, the hole-like FS was believed to be the only FS feature in
Bi2212 \cite{ding97,norman98,white96,onellion} over the doping
range going from underdoped to overdoped samples ($T_c\!\sim$15 K
and 67 K, respectively). These conclusions seem to be in contrast
with the case of LSCO, where a crossover from a hole to
electron-like FS is clearly evident near optimal doping
\cite{ino_dop,ino_dop1}. Recently, other reports questioned this
picture, arguing that one simple hole-like FS may not be a
complete characterization of the low-lying excitations in Bi2212
\cite{feng,chuang,gromko}. These studies suggested an
electron-like FS centered at the $\Gamma$-point
\cite{chuang,gromko} or, possibly, two co-existing electronic
components, resulting in electron and hole-like FSs \cite{feng}.
These suggestions were opposed by other groups which claimed that
only one hole-like FS is supported by the ARPES data once the
effects of the photon energy dependence of the matrix elements in
the ($\pi$,0) region are taken into account
\cite{fretwell,borisenko,mesot,bansil}.

\begin{figure}[t!]
\centerline{\epsfig{figure=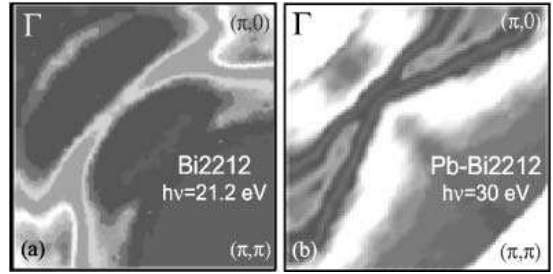,width=0.95\linewidth,clip=}}
\vspace{-.2cm} \caption{FS given by the integrated weight at $E_F$
for optimally (a) and Pb-doped (b) Bi2212 (from \cite{feng1} and
\cite{pasha0}).}\label{FS30meV}\end{figure}

To better illustrate the problems discussed above, we show in
Fig.\,\ref{FS30meV}a the FS of optimally doped Bi2212 determined
by integrating over a 7 meV window about $E_F$ the EDCs taken with
unpolarized HeI radiation (from \cite{feng1}). The data have been
measured on a whole quadrant of the BZ to verify the symmetry
between ($\pi$,0) and (0,$\pi$), and then have been symmetrized
with respect to the (0,0)-($\pi$,$\pi$) line to compensate for the
different data sampling along horizontal and vertical directions.
Inspection of the EDCs shows a Fermi crossing in going from
($\pi$,0) to ($\pi$,$\pi$), although the band is very flat
\cite{dessau93}, and from (0,0) to ($\pi$,$\pi$), in agreement
with the picture of a hole-like FS centered at ($\pi$,$\pi$). The
latter is observable in the $E_F$ intensity map of
Fig.\,\ref{FS30meV}a together with the two ghost FSs due to the
{\it umklapp} bands and located on each side of the primary FS. In
Fig.\,\ref{FS30meV}a we can also observe a considerable amount of
weight around ($\pi$,0). Because of the complexity of the EDCs in
this region and the absence of a clear drop in intensity along
(0,0)-($\pi$,0), one cannot make strong statements concerning an
electron-like FS.
\begin{figure}[t!]
\centerline{\epsfig{figure=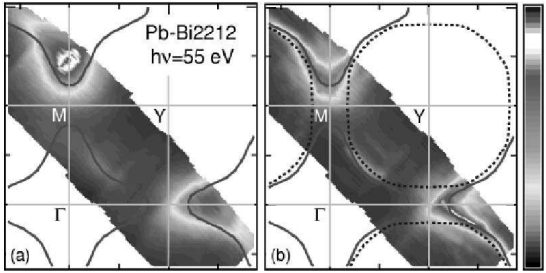,width=0.95\linewidth,clip=}}
\vspace{-.2cm} \caption{$n(k)$ plot (a), and $E_F$ intensity map
(b) obtained by integrating the EDCs over 580\,meV and 7\,meV
below $E_F$, respectively. Dotted blue lines indicate the
conventional hole-like FS; solid red lines the FS defined as the
locus of the leading-edge midpoints in the EDCs (from
\cite{pasha1}).}\label{Pasha_FS}\end{figure}

In relation to a possible electron-like topology of the Bi2212 FS
in the overdoped regime, interesting results have been very
recently reported by Bogdanov {\it et al.} \cite{pasha0,pasha1}.
The authors performed a detailed study of the FS topology in
slightly overdoped Pb-Bi2212 ($T_c$=82.5 K). The advantage of
working on Pb-doped Bi2212 is that Pb substitutes into the BiO
planes, suppressing the superstructure and therefore the
complications related to the {\it umklapp} bands at ($\pi$,0). As
shown by the $E_F$ mapping presented in Fig.\,\ref{FS30meV}b
\cite{pasha0}, this system represents a strong case for an
electron-like FS: in going from (0,0) to ($\pi$,0), a decrease of
intensity, corresponding to a Fermi crossing, is now clearly
observed near ($\pi$,0). However, as earlier data on the Pb-doped
system by Borisenko {\it et al.} \cite{borisenko} were interpreted
as evidence for a hole-like FS in Pb-Bi2212 also, and for the
universality of the hole-like FS topology in Bi2212, one must
examine these results in greater detail. Bogdanov {\it et al.}
\cite{pasha1}, used different methods and photon energies to
determine the FS in different BZs (Fig.\,\ref{Pasha_FS}). More
than 4000 EDC were collected and normalized by the integrated
intensity from a 100 meV window above $E_F$ (i.e., energy and
$k$-independent background), to compensate for variation of the
photon flux and for the non-uniform response of the detector. The
first indication for an electron-like FS was obtained from the
analysis of the leading-edge midpoint in the EDCs. The results
(independent of the zone) are plotted as solid red lines in
Fig.\,\ref{Pasha_FS}. This topology is confirmed by the $n(k)$
plot and $E_F$ intensity map (Fig.\,\ref{Pasha_FS}a
and\,\ref{Pasha_FS}b, respectively): both the maxima in the $E_F$
map and the steepest intensity drops in the $n(k)$ plot overlap
with the EDC-derived FS (red lines). Note that the intensity is
strongly suppressed in the first zone because of matrix element
effects. However, inspection of the EDCs shows that all the
features detected in the second BZ are present also in the first,
although significantly weaker.

As Pb doping does not modify the CuO$_2$ layer, the FS measured on
Bi2212 and Pb-Bi2212 is indicative of the electronic structure of
the same CuO$_2$ plane, and should show similar dependence upon
hole doping. While the results reported by Bogdanov {\it et al.}
\cite{pasha0,pasha1} do not rule out the coexistence of a
hole-like FS sheet, which may be apparent under different
experimental conditions \cite{fretwell,borisenko,mesot,bansil} as
shown, for example, in Fig.\,\ref{FS30meV}a \cite{feng1}, they
suggest that the scenario of a hole-like FS in Bi2212 over the
whole doping range is incomplete, and identify a possible
similarity between LSCO and Bi2212 along these lines.

Due to the limited doping range investigated, the ARPES results
from Bi2212 are insufficient to conclude in favor of any of the
four scenarios summarized in Fig.\,\ref{models}. While the FS of
optimally doped Bi2212 resembles the one reported in
Fig.\,\ref{models}a and\,\ref{models}d, in the underdoped region,
due to the opening of the pseudogap along the underlying FS (see
next section), the data is reminiscent of the models depicted in
Fig.\,\ref{models}b and\,\ref{models}c. The distinction between
Fig.\,\ref{models}a and\,\ref{models}d has to be determined on the
basis of the behavior near ($\pi$,0). This issue is probably
relevant to the controversy over the FS in the same $k$-space
region, and is important in connection to the possible signature
of the superfluid density in the ARPES data, as we will elaborate
later. The distinction between the models described in
Fig.\,\ref{models}b and\,\ref{models}c lies in the detection of
the `shadow FS' given by the dashed line in Fig.\,\ref{models}c.
Although it has been argued that the case of Fig.\,\ref{models}c
does not apply to Bi2212 \cite{ding97}, in contrast to an earlier
report \cite{aebi}, no consensus has been reached because, first,
the shadow FS may be simply too weak to be seen and, second, this
momentum region is complicated by the presence of shadow and {\it
umklapp} bands.

\subsubsection{Pseudogap}

 Despite the controversy over the topology of the FS,
many of the ARPES results are still reliable and the same holds
for the qualitative descriptions that were developed on the basis
of those results. This is particularly true with regard to a
doping dependent study performed under identical experimental
conditions. One of these important results is the normal state
excitation gap or pseudogap \cite{loeser,marshall96,dingN96}. This
feature was first recognized in the photoemission spectra in
attempting to connect the Bi2212 data to that from the insulator
\cite{king95,loeser,marshall96}. An example of the detection of
the pseudogap in ARPES spectra is presented in
Fig.\,\ref{Filip_DW}b where, following the idea of Laughlin
\cite{laughlin}, we compare the data at ($\pi$,0) as a function of
Dy concentration. As the Dy content is increased and the system
enters the underdoped regime, the `hump' shifts to high BE,
reflecting the opening of a normal state gap. As the size of the
pseudogap is rather large and the band dispersion near ($\pi$,0)
is weak, the detection of the pseudogap is relatively insensitive
to the FS topology: either FS topology will lead to qualitatively
similar conclusions. Hence, the normal state gap and its doping
dependence are robust features in the ARPES spectra.

The main characteristics of the pseudogap can be summarized as
follows \cite{hyperfine}: (i) the effect is strong in underdoped
samples, persists to optimal doping, and disappears in overdoped
samples \cite{loeser,marshall96,dingN96}. (ii) The gap has two
energy scales, with the low energy one being identified by the
location of the leading-edge midpoint, and the higher energy one
by the position of the hump. The lower energy scale has a $d$-wave
like momentum dependence, similar to the one of the
superconducting gap, but with a gapless arc near the nodal region.
The doping dependence of the two energy scales track each other
\cite{marshall96,norman98,white96,harris96,campuzano99}. (iii)
Upon decreasing the hole concentration, the size of the
leading-edge pseudogap increases, in contrast to the decreasing of
$T_c$. This provides an important piece of evidence for non-BCS
behavior of the superconducting transition in the underdoped
regime.
\begin{figure}[t!]
\centerline{\epsfig{figure=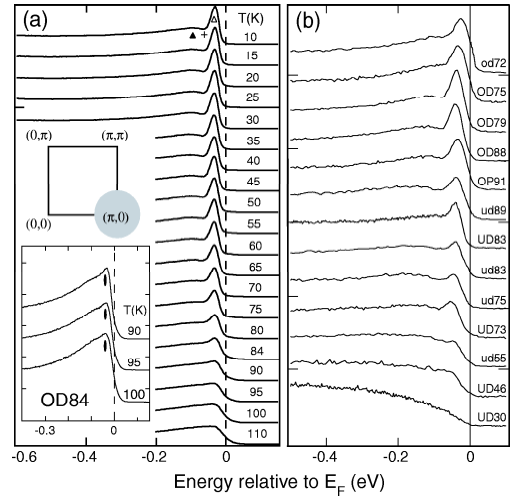,width=0.88\linewidth,clip=}}
\vspace{-.2cm} \caption{(a) $T$-dependent ARPES spectra from
overdoped Bi2212 ($T_c\!=\!84$ K). Data was collected around
($\pi$,0) in momentum space (shaded area in the sketch of 1/4-BZ).
Inset: enlarged view of EDCs taken just above $T_c$. (b) Doping
dependence of the superconducting-state ($\pi$,0) spectra of
Bi2212, for $T\!\ll\!T_c$. From \cite{feng3} and references
therein.}\label{DL_EDC}\end{figure}

\subsubsection{Signature of superfluid density}

We will now turn our attention to the well-known peak-dip-hump
feature \cite{zx,campuzano99,dessau91,hwu91,norman97} detected
below $T_c$ in Bi2212, in the region around ($\pi$,0). In
particular, we will discuss the detailed investigation of doping
and temperature dependence of this feature, which has been very
recently reported by Feng {\it et al.} \cite{feng3}. ARPES spectra
from Bi2212 were collected at many different temperatures in the
region around ($\pi$,0), as sketched in Fig.\,\ref{DL_EDC}a. Note
that this average in momentum space, which results in a
considerably improved signal-to-noise ratio, does not appreciably
alter the $k$-information because the QP dispersion is weak in
this region. In  Fig.\,\ref{DL_EDC}a, where EDCs from overdoped
Bi2212 with $T_c$=84 K (thus labeled as OD84) are displayed, we
can see the typical peak-dip-hump structure (open triangle, cross,
closed triangle, in Fig.\,\ref{DL_EDC}a). It becomes more
pronounced upon reducing the temperature below $T_c$ but it is
still visible slightly above $T_c$, as shown in the inset of
Fig.\,\ref{DL_EDC}a. The results obtained at low temperatures
($\sim\!10$ K) for different doping levels are displayed in
Fig.\,\ref{DL_EDC}b (UD for underdoped, OP for optimally doped,
and OD for overdoped). The peak, not observed in the very
underdoped samples, grows with doping and decreases again slightly
after optimal doping.

\begin{figure}[t!]
\centerline{\epsfig{figure=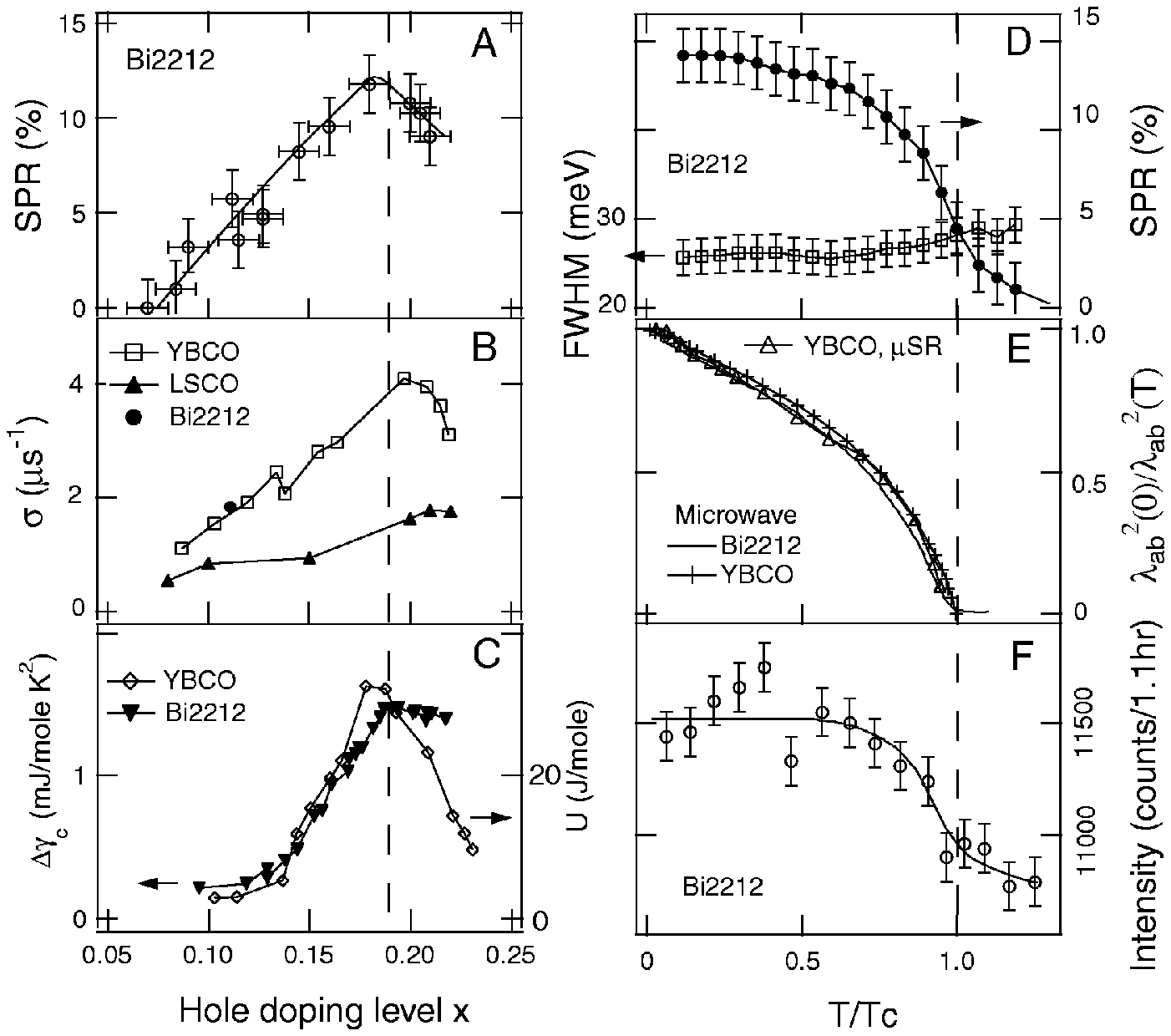,width=0.99\linewidth,clip=}}
\vspace{-.2cm}  \caption{Doping dependence, for $T\!<\!T_c$: (a)
SPR of Bi2212, from the spectra shown in Fig.\,\ref{DL_EDC}b
\cite{feng3}; (b) $\mu$SR relaxation rate ($\sigma\!\propto\!n_s$)
\cite{tallon99,uemura91}; (c) Bi2212 specific-heat coefficient
jump $\Delta\gamma_c\!=\!\gamma(T_c)\!-\!\gamma(120K)$
\cite{tallon00}, and YBCO condensation energy $U$ \cite{tallon99}.
$T$-dependence: (d) SPR and peak-width of Bi2212 (sample OD84 in
Fig.\,\ref{DL_EDC}a) \cite{feng3}; (e)
$\lambda_{ab}^2(0)/\lambda_{ab}^2(T)$ ($\propto\!n_s$)
\cite{jacobs,sonier99,bonn94}, where $\lambda_{ab}$ is the
in-plane penetration depth; (f) intensity of the neutron
($\pi$,$\pi$) mode \cite{he00}.}\label{DL_SPR}\end{figure}

Feng {\it et al.} \cite{feng3}, through a phenomenological fitting
procedure, were able to quantify the evolution of the peak
intensity. In order to extract meaningful and reliable information
(i.e., independent of artifacts due to $k$-dependence of matrix
elements and/or different experimental conditions for the
different samples), the authors focused on the ratio between the
relative intensity of the peak and the total spectrum intensity
(integrated from -0.5 to 0.1eV). The temperature and doping
dependence of this quantity, referred to as the `superconducting
peak ratio' (SPR), are presented in Fig.\,\ref{DL_SPR}a
and\,\ref{DL_SPR}d. From the comparison (Fig.\,\ref{DL_SPR}) with
many superfluid-related quantities measured in Bi2212,
YBa$_2$Cu$_3$O$_{7-\delta}$, and LSCO, two major conclusions can
be drawn: (i) the remarkable similarity of the data presented in
Fig.\,\ref{DL_SPR} strongly suggests a universality in the
superconducting properties of the cuprates; (ii) ARPES, which
mainly probes single-particle excitations of the condensate and
therefore directly measures the strength of the pairing (i.e.,
superconducting gap), can also provide information on the phase
coherence of the superconducting state (usually inferred from
those techniques which directly probe the collective motion of the
condensate).
This latter point is shown, as a function of hole concentration,
by the remarkable resemblance of the SPR to the superfluid density
($n_s$) measured by $\mu$SR (Fig.\,\ref{DL_SPR}b), the
condensation energy $U$ from the specific heat, and the jump in
the specific heat coefficient (Fig.\,\ref{DL_SPR}c). In addition,
upon increasing temperature the SPR decreases in a way similar to
$n_s$, as measured by microwave and $\mu$SR spectroscopy
(Fig.\,\ref{DL_SPR}d and\,\ref{DL_SPR}e), with an abrupt drop near
$T_c$ (disappearance of the phase coherence), rather than at $T^*$
(opening of the pseudo-gap in the underdoped regime).

As emphasized by Feng {\it et al.} \cite{feng3}, the sensitivity
of ARPES as well as of neutron experiments to the superconducting
condensate fraction \cite{sudip}, and the $x$ dependence of the
SPR and additional quantities in Fig.\,\ref{DL_SPR}, provide
direct evidence for the need of an approach beyond FL and BCS
theory. For instance, in treating the cuprates within the FL-BCS
framework, the coherence factor dictates that the QP spectral
weight depends on the magnitude of the energy-gap opened on the
normal-state FS, and the gap magnitude, as experimentally
determined by ARPES, scales as ($1\!-\!x$). To the contrary, the
SPR grows with $x$ in the underdoped region, and only in the
overdoped regime does the SPR scale as ($1\!-\!x$), possibly
indicating a crossover to a more conventional FL behavior.

As discussed by Feng {\it et al.} \cite{feng3}, there is no unique
interpretation for the observed phenomena. A possible explanation
for the $x$-dependence of the SPR can be provided by the stripe
model, similarly to what was discussed for LSCO. A calculation
\cite{carlson99} within this approach well reproduces the observed
T-dependence of the superfluid density, and identifies in the
weight of the superconducting peak a measure of the phase
coherence established in the 1D-2D crossover that, in the stripe
picture, accompanies the superconducting transition.
Alternatively, the nonmonotonic doping dependence of the SPR could
be a manifestation of a composite QP within the RVB approach
\cite{rvb,laughlin,affleck,wen,chakravarty,maekawa88,fukuyama88,kotliar88,ioffe96,anderson_sc,fisher,lee2000},
or of competing orders nearby a quantum critical point, where at
$T\!=\!0$ a quantum phase transition occurs, driven by quantum
rather than thermal fluctuations.

\subsubsection{Quasiparticle self energy}

Recently, detailed ARPES studies of the single particle self
energy $\Sigma(k,\omega)$  have been reported
\cite{norman_self,valla1,pasha2,kami}. These investigations are
extremely relevant because ARPES can measure both the real
(renormalized QP energy) and imaginary (inverse QP lifetime) parts
of $\Sigma(k,\omega)$, providing information on the many-body
character of the electronic excitations \cite{norman_self}. In the
following, we will in particular focus on the results obtained for
Bi2212 in the nodal region, where Valla {\it et al.} \cite{valla1}
found that the QP peak width shows a linear T-dependence
(independent of $\omega$, at small BE), and a linear
$\omega$-dependence (independent of T, at large BE). No change in
this behavior was found across $T_c$ (as may be expected on the
basis of DC resistivity), which excludes scattering-mediated decay
for the QP, contrary to the FL case. Valla {\it et al.}
\cite{valla1} argue that the observed scaling behavior is
suggestive of a nearby quantum critical point.

\begin{figure}[t!]
\centerline{\epsfig{figure=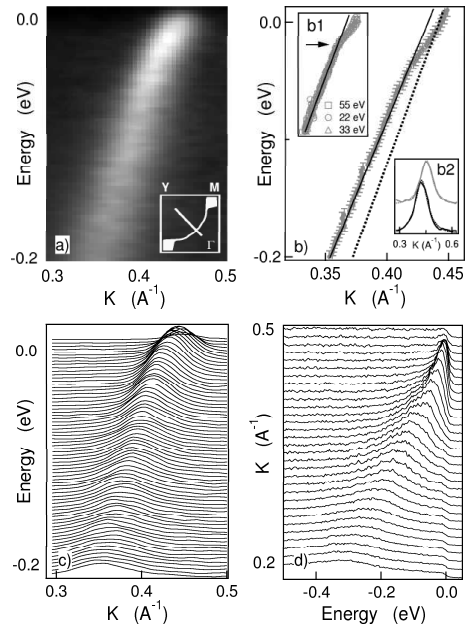,width=0.8\linewidth,clip=}}
\vspace{-.2cm} \caption{(a) Intensity plot $I(k,\omega)$ along
(0,0)-($\pi$,$\pi$) from slightly overdoped Bi2212 ($T_c$=91 K),
at 33 eV photon energy. MDCs and EDCs are shown in (c) and (d),
respectively. (b) QP dispersion, from fits of MDCs (inset shows
results for different photon energies). The dotted line is the LDA
dispersion. Inset (b2) MDCs (and correspondent fits) at 16 (black)
and 55 (gray) meV BE. Data from
\cite{pasha2}.}\label{Pasha_kink}\end{figure}

However, evidence for an additional energy scale in the QP self
energy was later reported by two other groups \cite{pasha2,kami}.
In particular, from the analysis of energy and momentum
distribution curves (MDCs), Bogdanov {\it et al.} \cite{pasha2}
found a kink at $50\pm15$ meV in the QP dispersion
(Fig.\,\ref{Pasha_kink}), contrary to the linear dispersion
predicted by LDA calculations. The kink appears to be at the same
BE along all directions in the BZ, and more pronounced at low
temperatures (with only a weak residual effect above $T_c$). In
addition, a drop in the low-energy scattering rate (related to the
kink) was found from the BE-dependence of the inverse QP lifetime.
Explanations proposed for this effect are electron-phonon
coupling, coupling to the neutron ($\pi$,$\pi$) mode, and stripes
\cite{pasha2,kami,norman00}. Among them the phonon scenario, which
is the most straightforward interpretation, might also be the most
plausible one. In fact, similar data has been obtained on systems
characterized by strong electron-phonon coupling
\cite{baer,valla2}.

At this stage no consensus has been reached on the above results
which have nonetheless opened a very promising direction in the
study of HTSCs. The information one can gain from the study of the
self energy (in particular, the presence or absence of extraneous
energy scales in the electronic self energy) is essential in
addressing the recent claim that the HTSCs are the realization of
a system where many-body correlations are so strong that the
electronic properties are not affected by conventional sources of
scattering (e.g., defects, impurities and thermal fluctuations)
which can only perturb the dynamics of individual particles
\cite{anderson_sc,laughlin00}.

\subsection{Nd$_{2-x}$Ce$_x$CuO$_4$}

As the final section of this discussion on the doping evolution of
the low energy electronic properties of the cuprates, we will
focus on the n-type superconductor NCCO. The relevance of a
comparative study of n and p-type HTSCs is that of a verification
of the symmetry, or lack thereof, between doping the AF insulator
with electrons or holes. This issue has important theoretical
implications because most models which are thought to capture the
essence of high-$T_c$ superconductivity implicitly assume
electron-hole symmetry \cite{peter}.
\begin{figure}[t!]
\centerline{\epsfig{figure=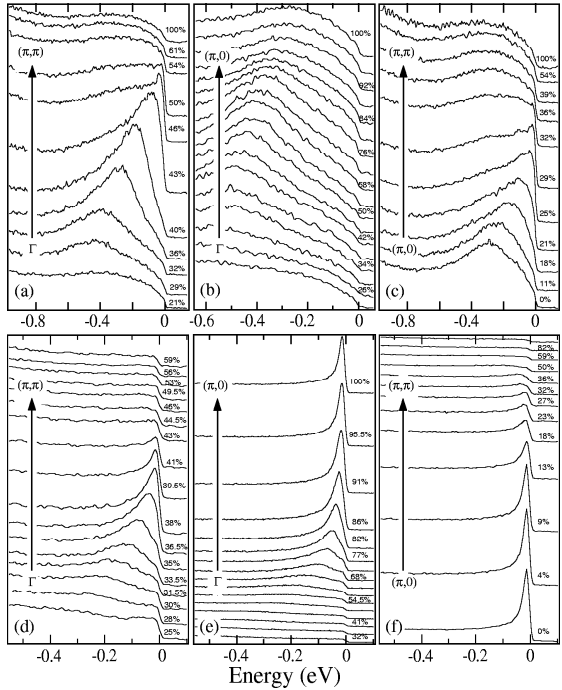,width=0.98\linewidth,clip=}}
\vspace{-.2cm} \caption{EDC along high symmetry lines for NCCO
(top panels) and Pb-Bi2201 (bottom panels), from \cite{peter}.
Data was collected at 10 K with 10-15 meV energy resolution and
momentum resolution of ~1\% of the BZ, and 16.5 eV and 21.2 eV
photons for NCCO and Pb-Bi2201, respectively.
}\label{Peter_EDC}\end{figure}

High energy and momentum resolution ARPES data on
Nd$_{1.85}$Ce$_{0.15}$CuO$_{4}$ has been recently reported by
Armitage {\it et al.} \cite{peter,peter1}. EDCs measured on NCCO
along the high symmetry lines in the BZ are shown in
Fig.\,\ref{Peter_EDC} (top panels) together with, for comparison
purposes, analogous spectra from single-plane slightly-overdoped
Bi$_{1.7}$Pb$_{0.3}$Sr$_{2}$CuO$_{6+\delta}$ (Pb-Bi2201), whose
behavior is generic to the p-type compounds (bottom panels). Along
the (0,0)-($\pi$,$\pi$) direction, NCCO and Pb-Bi2201 show a
dispersion which is ubiquitous among the cuprates, with a QP peak
dispersing quickly towards $E_F$ and crossing it at about
($\pi$/2,$\pi$/2). The exact crossing position varies with the
band filling, as shown also by the data in Fig.\,\ref{Peter_EDC}a
and\,\ref{Peter_EDC}d. Along (0,0)-($\pi$,0), however, the two
systems show a clear difference: while the low energy feature in
Pb-Bi2201 disperses close to $E_F$ and  around ($\pi$,0) forms a
flat band just below $E_F$, in NCCO the flat band is located at
300 meV below $E_F$ (Fig.\,\ref{Peter_EDC}b
and\,\ref{Peter_EDC}e).
\begin{figure}[t!]
\centerline{\epsfig{figure=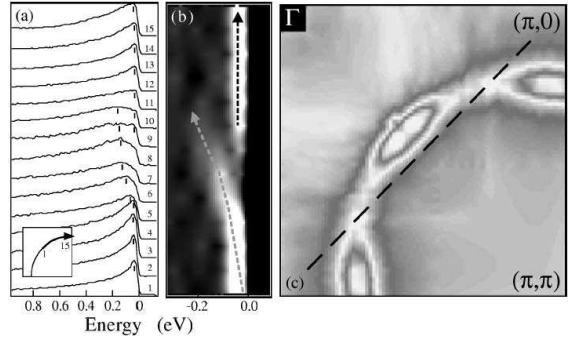,width=0.98\linewidth,clip=}}
\vspace{-.2cm} \caption{(a) EDCs of NCCO along the $k_F$ contour
[with the background defined as the signal at ($\pi$,$\pi$)
subtracted], from ($\pi$/2,$\pi$/2) to ($\pi$,0.3$\pi$) and,  (b),
their second derivative. (c) FS of NCCO obtained by integrating
over 30 meV at $E_F$ the EDCs from 1/8 of the BZ. The map was
symmetrized with respect to (0,0)-($\pi$,$\pi$) after direct check
of the symmetry along ($\pi$,0)-(0,$\pi$). The intensity
modulation along the FS highlights the intersection, at
(0.65$\pi$,0.35$\pi$) and (0.35$\pi$,0.65$\pi$), between the FS
and the {\it umklapp} surface represented by the dashed line. Data
taken from \cite{peter1}. }\label{Peter_FS}\end{figure}
Along ($\pi$,0)-($\pi$,$\pi$), the peak disperses to $E_F$ without
the same pronounced sharpening that occurs along the zone
diagonal. Moreover, in all spectra from momentum space points well
above the FS, a relatively large non-dispersive background
characterized by a broad maximum near 300 meV is observed.

In Fig.\,\ref{Peter_FS}c, the $E_F$ intensity map obtained by
integrating the EDCs in a 30 meV window about $E_F$ is shown
\cite{peter,peter1}. The location of the intensity maxima defines
a rounded piece of FS centered at ($\pi$,$\pi$), and characterized
by a volume greater than 1/2 (counting electrons) which is
consistent with Luttinger's theorem. For NCCO there is no
signature of the Fermi `patches' observed on the Bi-systems at
($\pi$,0) which are related to the presence of the large flat band
very close to $E_F$ (see, e.g., Fig.\,\ref{FS30meV}a which is
representative for both the 2212 and 2201 systems). On the other
hand, for NCCO distinct regions of reduced intensity are
identified along the FS contour near (0.65$\pi$,0.35$\pi$) and
(0.35$\pi$,0.65$\pi$).

The EDCs from along the FS [with the background defined as the
signal at ($\pi$,$\pi$) subtracted] and their second derivative
are presented in Fig.\,\ref{Peter_FS}a and\,b, respectively
\cite{peter1}. These spectra cannot be adequately explained by a
simple sharp peak at $E_F$ along the entire FS contour. At first
glance, they might appear to be describable in terms of two
separate features in the low energy electronic structure (marked
with ticks in Fig.\,\ref{Peter_FS}a and arrows in
Fig.\,\ref{Peter_FS}b). The feature at $E_F$ at ($\pi/2, \pi/2$)
appears to pull back towards higher binding energies as one moves
around the FS towards ($\pi$,$0.3\pi$).  As it moves to higher
energy it broadens becoming totally incoherent. Closer to ($\pi,
0.3\pi$) a second feature seems to appear at low energy while the
feature at high energy loses spectral weight and disappears. The
second derivative of the EDCs (Fig.\,\ref{Peter_FS}b) highlights a
discontinuity in intensity at those Fermi momenta where no sharp
QP peak is observed at $E_F$ in the EDCs. Armitage {\it et al.}
\cite{peter1} concluded in favor of an alternative description for
the lineshape of the EDCs in Fig.\,\ref{Peter_FS}a: a single
feature at $E_F$ which becomes incoherent at the intermediate
position. The spectral intensity is suppressed close to $E_F$
because much of the spectral weight is pushed to higher energies
\cite{peter1}. This would be the case, for instance, in the
scenario described in Fig.\,\ref{models}b, where the breakdown of
FL theory due to {\it umklapp} scattering is considered
\cite{rice}, with the difference that in the case of NCCO the FS
touches the {\it umklapp} surface (dashed line in
Fig.\,\ref{Peter_FS}a) at (0.65$\pi$,0.35$\pi$) and
(0.35$\pi$,0.65$\pi$), instead of ($\pi$,0) and (0,$\pi$) as in
Fig.\,\ref{models}b. This point of view that the two features are
part of a single spectral function is supported by the fact that
only a single FS with the expected Luttinger's volume is observed.

The results obtained on  NCCO by Armitage {\it et al.}
\cite{peter,peter1} seem to contrast with those from the p-type
HTSCs. The differences found in the spectral function of the n and
p-type materials seem to indicate that the electron-hole symmetry
usually assumed in the theoretical models describing the
low-energy properties of the CuO$_2$ plane, might be an
oversimplification. In particular, the local character of the
electronic states [doped electrons (holes) are thought to occupy
Cu (O) sites] which are eliminated by reducing the three bands of
the CuO$_2$ plane to one in the single-band Hubbard model, may
have to be reconsidered.

\section{Discussion and conclusions}

To date, the most complete ARPES studies are those on LSCO and
Bi2212, because these materials can be investigated over a wide
doping range. The LSCO system has been extensively interpreted in
the stripe scenario because it provides a possible explanation for
many of the experimentally observed features: (i) the
two-component electronic structure seen in the very underdoped
regime that is suggestive of the creation of new electronic states
inside the Mott gap; (ii) the lack of chemical potential shift in
the underdoped regime; (iii) the straight FS segments observed
under certain experimental geometries that are indicative of 1D
electronic behavior and cannot be reconciled with the band-like FS
(Fig.\,\ref{models}a). What is also interesting is that the LSCO
results seem to suggest a `dual nature' for the low-lying
electronic excitations \cite{zhou1}: on the one hand, the data
shows the effect of charge ordering as mentioned above. On the
other hand, possibly as a consequence of fluctuations, disorder,
or incomplete charge disproportionation, it exhibits features that
deviate from the simple rigid stripe picture, such as the nodal QP
weight and the QP dispersion along directions perpendicular to the
stripes. Furthermore,the FS becomes better defined upon increasing
doping, and changes from hole-like to electron-like near optimal
doping. In Nd-LSCO for a given doping, it also becomes better
defined upon decreasing the Nd concentration whose main role is to
pin stripe fluctuation. All these observations seem to suggest
that, as doping is increased, LSCO becomes more of a band-like
system with an LDA-like FS, and the signatures of stripes weaken;
near optimal doping, the distinction between the pictures reported
in Fig.\,\ref{models}a and\,\ref{models}d would become blurred. As
shown recently by Mizokawa {\it et al.} \cite{mizokawa}, the dual
electronic structure would also indicate that the electronic
properties of a system in the stripe phase are different from
those of a truly 1D-chain system, in that the transverse motion is
allowed in the stripe case but not in the 1D-chain case.

For the Bi2212 system, on the other hand, there is no extensive
and reliable data in the underdoped regime and, in particular,
near the MIT boundary. Therefore, those features which have been
discussed as possible signatures of a charge ordered state in
extremely underdoped LSCO, have not been confirmed so far in the
case of Bi2212. However, many other features seen on Bi2212 are
very similar to those of LSCO. These include the doping dependence
of the ($\pi$,0) pseudogap, with the leading-edge gap being
smaller in LSCO; possibly, the evolution of the FS from hole-like
to electron-like with doping [although the results obtained around
the more complicated ($\pi$,0) region are still controversial];
the kink in the dispersion of the QP, which has recently been
observed in LSCO and Nd-LSCO as well \cite{zhou1}. In the case of
Bi2212, it seems that all the models summarized in Fig.
\ref{models} are capable of providing a self-consistent, albeit
not unique interpretation. Although the normal state spectra of
this material in the underdoped and optimally doped cases are very
broad, one can still define a FS that becomes better defined at
higher doping levels (see Fig.\,\ref{models}a
or\,Fig.\,\ref{models}b, respectively, for optimally and
underdoped regime). On the other hand, the doping dependence of
the superconducting peak near ($\pi$,0), where the SPR scales as
$x$ contrary to what is expected within the FL approach, favors
the models depicted in Fig.\,\ref{models}c and\,\ref{models}d,
which include the essential ingredients of the physics of the
doped Mott insulator.

At this stage of the research in the HTSCs and their undoped
parent compounds, it does not seem possible to firmly conclude in
favor of one particular comprehensive theoretical model, in spite
of the considerable progress made in recent years. This situation
is exemplified by the longstanding puzzle concerning a fundamental
question, i.e., how does the doping of a Mott insulator take place
(Fig.\,\ref{mott}). On the one hand, very recent experimental
results  \cite{sawa_pri,filip_pri} favor a scenario based on the
shift of the chemical potential to the top of the LHB (or to the
bottom of the UHB for the n-type systems). This is in agreement
with the results from the $t$-$t'$-$t''$-$J$ model for SCOC, which
reproduce the substantial deformation of the QP band structure
upon doping, and suggest a unifying point of view for both the
undoped insulator and the HTSC \cite{eder}. On the other hand, in
the case of LSCO, the lack of chemical potential shift observed
in the underdoped regime \cite{ino_mu} and the detection of
multiple electronic components \cite{ino_nodal} support the
formation of in-gap states upon doping the system and,
consequently, the need for a completely new approach. In order to
establish whether the evolution from the Mott insulator to the
HTSC is truly accounted for  by one of the existing models, or
whether a completely different approach is required, these points
need to be further clarified.

The results discussed in the course of this review demonstrate
that the cuprates are complex materials characterized by many
competing degrees of freedom, which lead to different peculiar
physical properties coexisting with $d$-wave superconductivity.
The signatures of many-body effects seen in the behavior of the
superconducting peak detected on Bi2212 at ($\pi$,0), and the kink
in the QP dispersion observed in both Bi2212 and LSCO are
providing new experimental clues, whose theoretical importance
will likely emerge in the time to come.

\section{Acknowledgements}

We acknowledge N.P. Armitage, P. Bogdanov, D.L. Feng, Z. Hussain,
S.A. Kellar, C. Kim, A. Lanzara, F. Ronning, K.M. Shen, T.
Yoshida, and X.J. Zhou for discussions and many useful comments.
SSRL is operated by the DOE office of Basic Energy Research,
Division of Chemical Sciences. The office's division of Material
Science provided support for this research. The Stanford work is
also supported by NSF grant DMR9705210 and ONR grant
N00014-98-1-0195.

\end{document}